# Chiral Quasicrystalline Order and Dodecahedral Geometry in Exceptional Families of Viruses

O.V. Konevtsova[1,2], S.B. Rochal[1,2] and V.L. Lorman[2]

[1]*Faculty of Physics, Southern Federal University, 5 Zorge str., 344090 Rostov-on-Don, Russia*
[2]*Laboratoire Charles Coulomb, UMR 5221 CNRS and Université Montpellier 2, pl. E. Bataillon, 34095 Montpellier, France*

On the example of exceptional families of viruses we i) show the existence of a completely new type of matter organization in nanoparticles, in which the regions with a *chiral* pentagonal *quasicrystalline order* of protein positions are arranged in a structure *commensurate* with the spherical topology *and dodecahedral geometry,* ii) generalize the classical theory of quasicrystals (QCs) to explain this organization, and iii) establish the relation between local chiral QC order and nonzero curvature of the dodecahedral capsid faces.

The entry of a virus into a host cell depends strongly on the protein arrangement in a capsid [1], a solid shell composed of identical proteins, which protects the viral genome from external aggressions. In spite of a certain similarity in organization between capsid structures and classical crystals, the generalization of solid state physics concepts taking into account specific properties of proteins has started only quite recently [2]. Almost all works devoted to the physics of viruses with spherical topology postulate as an initial paradigm the existence of a local *hexagonal order* of protein positions [3,4]. Indeed, a large number of spherical viruses with global icosahedral symmetry show this type of organization. Cryomicroscopy data confirming this fact [5] gave rise to capsid structure models based on the relation between the order in a plane hexagonal lattice and the geometrical properties of an icosahedron (or more complex polyhedra belonging to the icosahedron family). Pioneering models in the field use the commensurability between the plane hexagonal lattice and the plane nets of these polyhedra [3]. Among the direct consequences of the local hexagonal order hypothesis there are i) drastic selection rules limiting the number of proteins and their environments in a capsid [3,4], and ii) location of topological defects of the hexagonal (or hexatic) order in the vicinity of fivefold axes of the global icosahedral symmetry. The latter property has allowed the authors of Ref. [6] to argue that faceting of large viruses is caused by a buckling transition associated with the 12 defects of fivefold symmetry [7]. However, there exist exceptional families of viruses (e.g., papovaviruses of papilloma and polyoma families) which show a local *pentagonal order* of capsid proteins [see Fig. 1(a)] and not a hexagonal one [8]. For these families, the principles of organization, defect formation, and mechanical properties (including faceting) remain uncovered.

The present Letter aims to give a clear-cut explanation of the unusual pentagonal protein arrangement in viral capsids from the viewpoint of solid state physics, namely, quasicrystal (QC) theory. We show that the systems considered represent the first example of matter organization in nanoparticles, in which the regions with a *chiral* pentagonal *quasicrystalline order* of protein positions are arranged in a structure *commensurate* with the spherical topology *and dodecahedral geometry;* we establish the relations between the elastic properties of these unconventional QC regions and the unusual structure and geometry of the resulting capsid.

The capsids under consideration are constituted by $60M$ proteins, where $M$=6, the number which is forbidden by the classical selection rules [3]. In addition, the structures show no fivefold defects. Pentagonal order being perfectly compatible with global fivefold symmetry axes of the capsid, its topological defects find themselves in *20 isolated points of threefold symmetry* instead. This fact suggests that structural peculiarities of these capsids correspond to a *dodecahedron geometry* and not an icosahedron one. Consequently, their mechanical properties and faceting mechanisms are also related to the dodecahedron geometry.

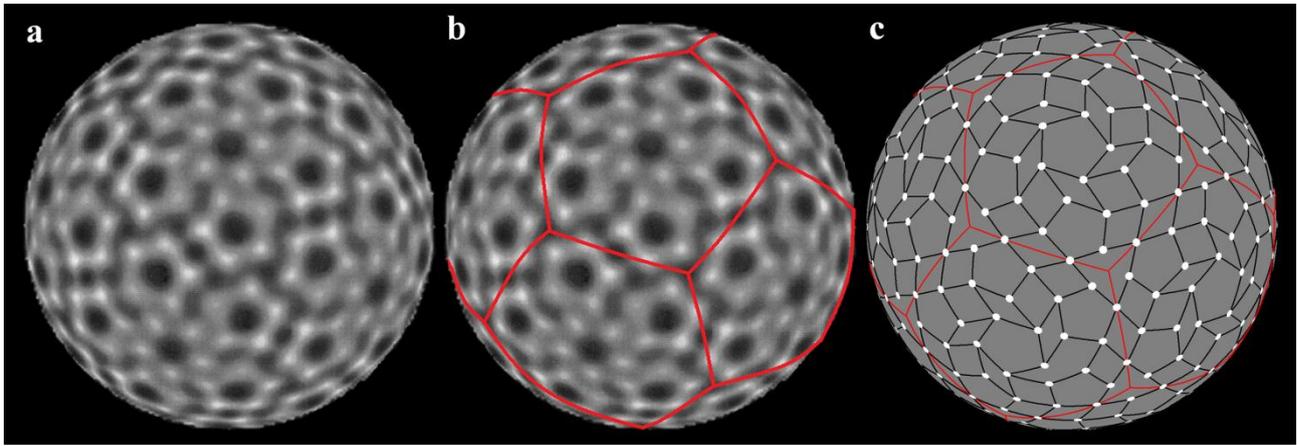

FIG. 1 (color online). Protein locations in a capsid of bovine papilloma virus. (a) Experimental protein density map. (b) Superimposition of the protein density map with a dodecahedral tessellation of the sphere. (c) The idealized quasilattice of protein density maxima.

Figure 1(a) shows an experimental protein density map in a capsid of bovine papilloma virus (BPV) [8]. The maxima of intensity (centers of clear circles) correspond to positions of protein centers. The BPV density distribution is not compatible with the tessellation of a sphere in 20 spherical triangles, usually used in structural virology to classify protein distributions with hexagonal order. Instead, we divide a sphere into 12 spherical pentagons [Fig. 1(b)]. The vertices of the tessellation obtained coincide with the vertices of the *dodecahedron* inscribed in the sphere. In Fig. 1(c) the protein center positions are presented on a spherical surface and connected by lines of approximately equal length. Being projected on the dodecahedron surface, protein positions occupy the nodes of extremely regular plane tiling with local pentagonal symmetry. All edges connecting the nodes of the plane tiling become of exactly equal length. The dodecahedron faces are decorated with the same tiles (equal regular pentagons, thin and thick rhombuses) and in exactly the *same way*. These properties make proteins in the capsid *quasiequivalent* [3], and, consequently, minimize the number of different protein conformations necessary for the capsid self-assembly (see Appendix I). The *defects* of the pentagonal order form equilateral triangles with the centers located *in the dodecahedron vertices* [Fig. 1]. This interpretation is in sharp contrast with all existing approaches (including the approach in terms of tilings developed in [10]) to the BPV structure. Nevertheless, a similar analysis that we performed in a whole series of papovaviruses (experimental density distributions were taken from [11]) confirmed that dodecahedral geometry (and not only pentagonal order) is common for capsids of this family.

Indeed, in each pentagonal face of the dodecahedron the nodes occupied by proteins belong to the same plane pentagonal quasilattice [12] (though the number of nodes is limited by the face size): the positions of neighboring nodes are transformed one into another by one of the *five orientationally equivalent 2D translations* (edges of the tiles) which are parallel to the edges of the face (see Fig. 2). This quasilattice shares a series of common features with the classical QC in metal alloys [13] or more recently discovered QC in micellar [14] or polymer systems [15]. However, viral "nanoquasicrystals" have three striking peculiarities. First, the symmetry axes of the viral quasilattice are not located in its nodes. Thus, the *nodes have no proper symmetry* and can be occupied by completely asymmetric proteins. This point is crucial for the structures composed of proteins, the asymmetry being one of the main structural characteristics of these biomolecules. Second, the arrangement of nodes is *chiral*, and thus is also compatible with the asymmetry of individual proteins. Both properties have no analog in QC systems. And third, finite pieces of 12 2D quasilattices are arranged in a structure commensurate with the dodecahedral geometry.

Among all known plane pentagonal quasilattices there exists only one [see Fig 2(a)], the so-called pentagonal Penrose quasilattice [12,16] (PPQ) with the organization of nodes similar to that shown in Fig. 1(c). The PPQ differs from the classical Penrose tiling consisting of two types of rhombuses. Its symmetry axes are not located at its nodes. Therefore, the PPQ is suitable for a decoration by asymmetric chiral proteins. The relation between the PPQ and the experimentally observed protein distribution is shown in Figs. 2(b) and 2(c). It can be presented as a two-step switching of several occupied positions. Each step concerns only one orbit of the $C_{5v}$ symmetry group of the plane quasilattice. Such kind of switching is well known in classical QCs [17] and is usually called phason jumps.

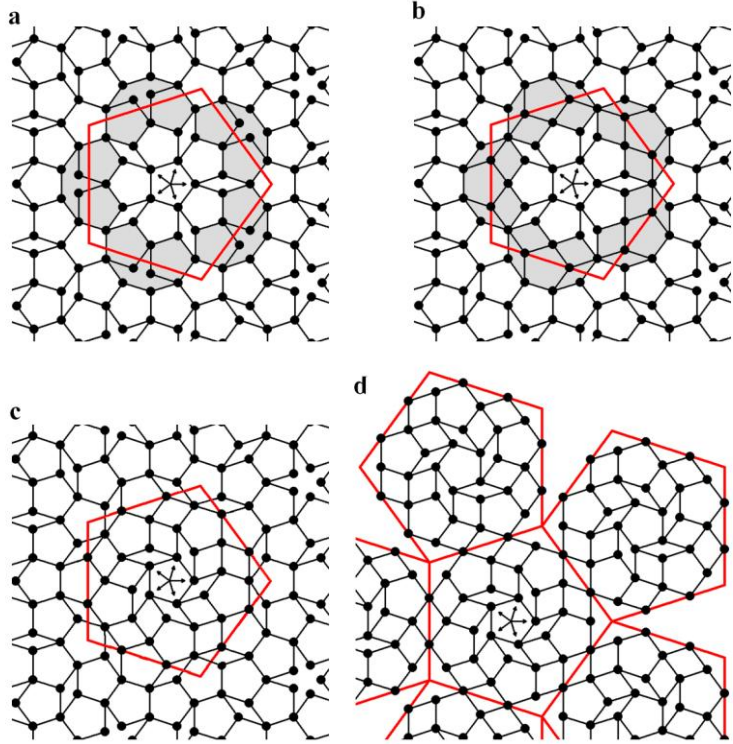

FIG. 2 (color online). Pentagonal QC order commensurate with the dodecahedron 2D net and the BPV capsid. Protein positions are presented by full circles. Dodecahedron faces are shown by large red pentagons. Five arrows in the center of each panel correspond to 2D basis translation vectors $\mathbf{a}_i$ of the quasilattice. (a) Conventional pentagonal Penrose quasilattice (PPQ). (b) Quasilattice after the action of the phason strain field responsible for the matching of the pentagonal order at dodecahedron faces. The resulting switching of nodes is shown in the shaded zone. (c) The final form of the quasilattice after the action of the chiral phason strain field. (d) The dodecahedron 2D net decorated with the chiral pentagonal quasilattice (six of 12 faces of the net are shown). In the folded form it corresponds to the BPV capsid.

Physically, the phason jumps in PPQ resulting in the viral capsid QC come from the unconventional dodecahedral geometry of the capsid. First, in viral QCs the plane pentagonal structure in each dodecahedron face admits the possibility of exact matching with adjacent faces at the dodecahedron edges. The matching in the vicinity of face boundaries induces a slight reconstruction of the QC order preserving the quasilattice symmetry. Second, in a real virus capsid the local curvature is different from zero, the faces of the "dodecahedron" are *curved,* thus changing distances between the nodes of the quasilattice. We show below that the resulting in-plane strain induces an additional *chiral* switching of the PPQ nodes. These two steps of switching are described in terms of the QC physics as a single *low-energy nonlinear phason strain field*. To deduce this field we minimize the elastic free energy of a dodecahedron face with the boundary conditions which take into account the matching.

For the perfect matching, the pentagonal faces of the 2D net of the dodecahedron should contain twofold axes in the middle of each edge [see Fig. 2(d)]. Introduction of twofold axes leads to the simple nonlinear phason strain $\mathbf{v}^0 = \mathbf{v}^0(\mathbf{r})$, preserving the symmetry of the unstrained quasilattice. This results in a correlated switching of ten nodes closest to the dodecahedron face boundaries [see the shaded zone in Figs. 2(a) and 2(b)]. Note that, in terms of the 5D space $E$ (Appendix II), the twofold axes which appear in the net are related to the superposition of the inversion and a translation from the star <5,5,-1,-2,-1>. This translation is the sum of 5D coordinates of two positions lying at the face boundary in Fig. 2(b). Applying group theory analysis we obtain the functional form of the phason strain field $\mathbf{v}^0$ responsible for the matching. Indeed, the field $\mathbf{v}^0 = (v_x^0, v_y^0)$ and the radius vector $\mathbf{r} = (x, y)$ span two different representations of the $C_{5v}$ symmetry group, the former representation being contained in the symmetric square of the latter. To preserve fivefold symmetry in each face the field $\mathbf{v}^0$ satisfies the condition $\mathbf{v}^0(\mathbf{0}) = \mathbf{0}$. Then, the explicit form of the *phason strain field responsible for the matching* is:

$$v_x^0 = \alpha(y^2 - x^2) \; ; \quad v_y^0 = 2\alpha xy \qquad (1)$$

Parameter $\alpha$ is a constant or an arbitrary function of the distance $r = \sqrt{x^2 + y^2}$ from the face center. Already in the case $\alpha = const$ there exists an extended $\alpha$ region that corresponds to the perfect matching of pentagonal order at the dodecahedron edges.

Equations (1) can be easily justified energetically in the frame of the elasticity theory of QCs. The harmonic elastic energy density of a pentagonal QC [18] contains two invariants composed of the first derivatives of an arbitrary phason strain field $\mathbf{v}(\mathbf{r})$: $J_1 = (\partial_x v_x)^2 + (\partial_y v_x)^2 + (\partial_x v_y)^2 + (\partial_y v_y)^2$ and $J_2 = (\partial_x v_x)(\partial_y v_y) + (\partial_y v_x)(\partial_x v_y)$. In each face of the dodecahedron the elastic energy functional

$$F = \int_S \left( K_1 J_1(\mathbf{v}) + K_2 J_2(\mathbf{v}) \right) dS \qquad (2)$$

is minimized with the boundary conditions making the matching at the edges possible. Namely, the integral over the boundary of the scalar product $(\mathbf{v}\mathbf{v}^0)$ should be equal to the integral over the same boundary of $(\mathbf{v}^0)^2$, where $\mathbf{v}^0$ is field (1). For the sake of simplicity, the pentagonal face $S$ is replaced by a circle containing the same nodes of the quasilattice. Here $K_1$ and $K_2$ are phason elastic constants of the pentagonal QC. This variational problem has an exact solution. It is remarkable that its form is given by Eq. (1), which was obtained above using group theory arguments.

To account for the remaining peculiarities of papovavirus nanoquasicrystals we consider the coupling between the *chiral component of phason strain* and the in-plane conventional (phonon) strain, which is possible *in chiral quasilattices* only. In the chiral pentagonal QC the elastic free energy density contains an additional (with respect to the case of achiral QC) term. It is a pseudoscalar term with respect to the symmetry $C_{5v}$, but it becomes invariant in a chiral QC with the symmetry $C_5$. This situation is well known, for example, in the theory of chiral liquid crystals [19,20], where an additional pseudoscalar term ($\mathbf{n}$ curl $\mathbf{n}$) linear in first derivatives of the director $\mathbf{n}$ appears in the Frank-Oseen elastic free energy of a cholesteric. However, in chiral pentagonal QCs there are no purely phason terms of this type. Indeed, $C_5$ symmetry forbids the existence of invariant terms composed only of the first derivatives of the field $\mathbf{v}$. By contrast, chiral symmetry allows the term linear with respect to the first derivatives of both $\mathbf{u}$ and $\mathbf{v}$ fields:

$$J_3 = (\varepsilon_{xx} - \varepsilon_{yy})(\partial_y v_x - \partial_x v_y) + 2\varepsilon_{xy}(\partial_x v_x - \partial_y v_y), \qquad (3)$$

where $\varepsilon_{ij}$ are the components of the usual (phonon) plane strain tensor. This term was discussed previously in the context of dislocation theory in-plane QCs [18, 21]. Because of the coupling (3) the inhomogeneous in-plane strain resulting from the capsid faces' curvature induces the chiral phason strain in the system. We suppose here that the deviation from the flat face geometry has approximately radial character. Then, the coordinates $\mathbf{R}$ of a point on a spherical segment of the radius $R$ are related to the coordinates of its projection $<x,y,h>$ on the flat face situated at the distance $h$ from the center along the $z$ axis as:

$$\mathbf{R} = \frac{R}{\sqrt{x^2 + y^2 + h^2}} <x, y, h>, \qquad (4)$$

The corresponding strain tensor is expressed as $\boldsymbol{\varepsilon} = (\mathbf{M}_s - \mathbf{M}_p)/2$ [20], where

$$\mathbf{M}_s = \frac{R^2}{(x^2+y^2+h^2)^2}\begin{bmatrix} y^2+h^2 & xy \\ xy & x^2+h^2 \end{bmatrix} \quad (5)$$

is the metric tensor of the spherical segment, and $\mathbf{M}_p$ is the metric tensor of the plane (i.e., unit matrix). The explicit form of the coupling term in the elastic energy density is obtained by substituting the strain tensor in Eq. (3). In the following consideration we suppose that the deviation from the flat face geometry is weak: this corresponds to the experimental faceted shape of the BPV capsid [8]. This means that $R>>R_{cap}$, where $R_{cap}$ is the capsid radius, or equivalently $h \approx R$. Then, expanding the coupling term (3) in a series and keeping the first nonvanishing terms in $<x, y>$ we obtain:

$$J_1 = \frac{(x^2-y^2)(\partial_y v_x - \partial_x v_y) + 2xy(\partial_x v_x - \partial_y v_y)}{2R^2} \quad (6)$$

The elastic free energy which takes into account all the peculiarities of the chiral QC order in the viral capsid is expressed in the form:

$$F = \int_S \left(K_1 J_1(\mathbf{v}) + K_2 J_2(\mathbf{v}) + K_3 J_3(\mathbf{v})\right) dS \quad (7)$$

The energy (7) is then minimized with the boundary conditions at the glued edges of the dodecahedron identical to those described above for energy (2). Here $K_3$ is the elastic constant of the *chiral phason-phonon coupling*. Even in this complex case the variational problem has an exact solution:

$$v_1^0 = \alpha(y^2 - x^2) + \beta(y^3/3 - x^2 y) \; ; \quad v_2^0 = 2\alpha xy + \beta(x^3/3 - x y^2) \quad (8)$$

Equation (8) expresses the *total phason field* in the virus "nanoquasicrystal." The chiral component of field (8) depends on the parameter $\beta$. The value of $\beta$ is determined, in turn, by the values of the elastic constants $K_1$, $K_2$ and $K_3$, and by the radius R: $\beta = K_3/[6R^2(2K_1 + K_2)]$. The corresponding minimal free energy per one dodecahedron face is expressed then as: $F_0 = r_0^4 \pi [96 R^4 \alpha^2 K_1^2 - 24 R^4 \alpha^2 K_2^2 - r_0^2 K_3^2]/[12 R^4 (2K_1 + K_2)]$, where $r_0$ is the effective radius of the face. Equation (8) together with the PPQ construction algorithm (see Appendix II). give the explicit procedure for determining the protein positions in the final chiral structure. In the extended region of $\alpha$ and $\beta$ values, the calculated structure perfectly corresponds to the experimental protein density distribution in BPV [Fig. 1]. For instance, for $\alpha \approx 0.03$ the value of $\beta$ is in the interval [0.0015-0.012]; for $\alpha \approx 0.024$ the value of the chiral field component is about $\beta \approx 0.009$. The resulting chiral structure is shown in Figs. 2(c) and 2(d).

In conclusion, we have shown that proteins in the BPV (and other viruses of the exceptional papovavirus families) self-assemble into unprecedented chiral QC-like structures, with chiral pentagonal order in the faces and global dodecahedron geometry of the capsid. The corresponding dodecahedron net is commensurate with the chiral pentagonal quasilattice and is tiled in the way presented in Figs. 2(c) and (d), with the asymmetric proteins put at the nodes of the quasilattice. We developed the nonlinear phason strain concept in the frame of the classical elasticity theory of QCs. The resulting approach allowed us to calculate the protein positions and thus to explain the protein organization in papovaviruses, in spite of its extreme complexity. The results obtained in this work constitute the fundamental basis for further studies of the exceptional virus self-assembly thermodynamics, mechanics of the capsids with pentagonal QC order, and mechanisms of their faceting.

# Appendix I. Exceptional capsids and quasiequivalence principle introduced by Caspar and Klug

The quasiequivalence principle was introduced by Caspar and Klug (CK) [3] for polyhedral viral capsids. According to this principle, proteins in a capsid should be located in quasiequivalent positions in order to minimize the number of different protein conformations necessary for the capsid self-assembly. CK proposed also a geometrical model for viral capsids. In this model, the net of a polyhedron of the icosahedron type is cut in a plane periodic hexagonal structure containing no inversion nor symmetry planes. The CK geometrical model satisfies perfectly the quasiequivalence principle and it is not surprising that the majority of viral capsids are organized in accordance with this model. However, the quasiequivalence principle can be satisfied in geometrical models different from the CK one. This point becomes crucial for the capsids with the number of proteins incompatible with the CK model (it is the case of 360 proteins in capsid of papovaviruses considered in the present work). In the model based on the PPQ, proteins are located in the nodes of the quasilattice, thus decreasing the number of protein conformations, analogously to the case of the periodic structure proposed by CK. Proteins in the PPQ model are bound to their neighbors in a rather uniform way and, consequently, satisfy the quasiequivalence principle, though the degree of quasiequivalence is lower than that in capsids compatible with the CK model.

The quasiequivalence principle helps to explain the fact that capsids with QC organization containing more than 72 pentamers do not exist in nature. Indeed, in capsids with 72 pentamers, proteins form three very symmetric types of tiles: pentagons and two types of rhombuses. However, the infinite PPQ contains, in addition, other types of tiles, namely stars and truncated stars. For quasicrystalline capsids containing more than 360 proteins these star-like tiles will necessarily appear on the polyhedral capsid surface, thus decreasing the degree of quasiequivalence. The self-assembly of the corresponding QC capsids should involve a much greater number of protein conformations. This fact makes the self-assembly of big QC capsids quite difficult.

Let us finally note that the QC capsid structure presented in the main part of the present work [Fig. 1(c) and 2(d)] is the only structure composed of 72 pentamers which corresponds to the minima of the elastic energy of a QC with the boundary conditions imposed by the dodecahedron geometry. However, a simple (but rigorous) geometrical analysis shows that there exist three other "mutant" capsids composed of 72 pentamers which cannot be obtained by the QC free energy minimization, though they satisfy the same geometrical principals as the capsid of BPV does. The geometrical principals put in the basis of this analysis are the following: 1) Any protein in the capsid belongs to one of the pentamers (there are no proteins which do not belong to a pentamer, and, there is no protein which belongs to more than one pentamer); 2) Everywhere on the capsid surface, except at dodecahedron vertices, the pentamers are conjugated by thin and thick rhombuses with the acute angles $\pi/5$ and $2\pi/5$, respectively. These principals are consistent with the geometrical model of the QC organization of the pentagonal capsid faces developed in the main part of the work. Three "mutant" structures obtained differ only slightly from the structure minimizing the free energy [Fig. 1(c) and 2(d)]. The difference resides in the organization of these capsids in the vicinity of dodecahedron edges only. The nets of the "mutant" capsids which do not correspond to minima of the free energy functional (Eq. (7) in the main part of the work) are presented in the following figure.

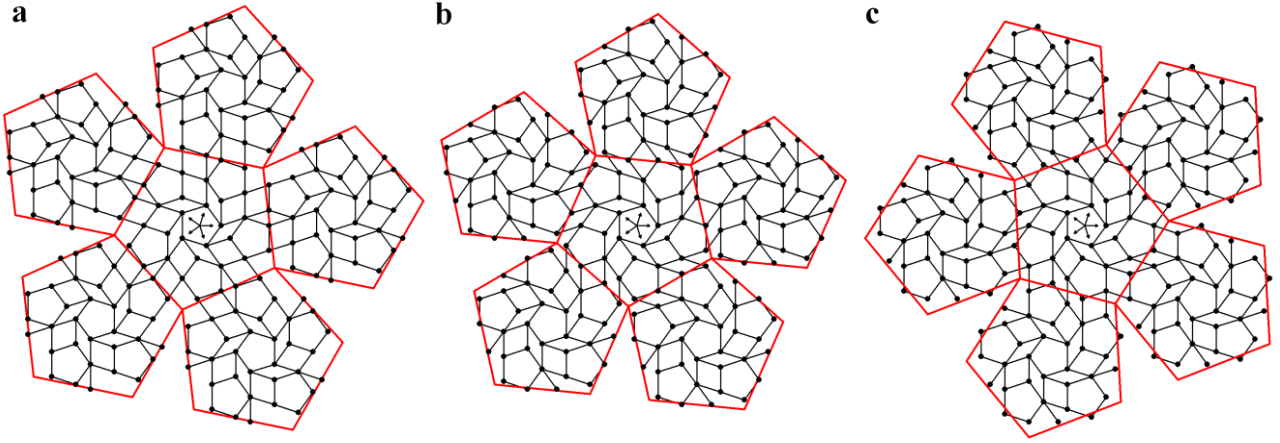

FIG. S1. "Mutant" capsids which do not correspond to the free energy minima: three additional possible dodecahedral nets of a capsid with all-pentamer QC organization of 360 proteins (six of twelve faces of the net are shown). Protein positions are presented by full circles. Dodecahedron faces are shown by big red pentagons. Five arrows in the center of each panel correspond to 2D basis translation vectors $\mathbf{a}_i$ of the quasilattice. Two-fold axes of the net transforming dodecahedron faces one into another correspond in the 5D space $E$ to superpositions of the inversion with a 5D translation. For the structures presented in (a-c) these translations are generated by cyclic permutations of the components of following vectors: (a) <2,-1,-3,2,6>; (b) <3,-1,-3,2,5>, and (c) <4,5,0,-3,0>.

## Appendix II. Pentagonal Penrose Quasilattice

The description of a pentagonal Penrose quasilattice (PPQ) is based on the projection from a 5D space $E$. The integer coordinates $\{n_i^j\}$ (with i=0,1…4) of a $j$-th point in $E$ have three irreducible projections spanning three irreducible representations of the 2D pentagonal symmetry group $C_{5v}$. Two different vector representations of $C_{5v}$ define the projections onto so-called direct and perpendicular 2D spaces [12,17]. The third one is identity representation. The value of its basis function $\sum_{i=0}^{4} n_i^j = \xi$ defines similar but slightly different pentagonal quasilattices. The coordinates of PPQ node $j$ are defined as:

$$\mathbf{r}^j = \sum_{i=0}^{4} n_i^j \mathbf{a}_i, \qquad (1)$$

where $\mathbf{a}_i$ are 2D basis translation vectors chosen in the following form: $\mathbf{a}_i = (\cos(i 2\pi/5), \sin(i 2\pi/5))$ with $i=0,1,2,3,4$. In Fig. 2 these $\mathbf{a}_i$ vectors are shown as the five arrows in the center of each panel. The node belongs to the quasilattice if its coordinates in the perpendicular space

$$\mathbf{r}_j^\perp = \sum_{i=0}^{4} n_i^j \mathbf{a}_i^\perp + \mathbf{v}, \qquad (2)$$

(where $\mathbf{v}$ is the phason field; $\mathbf{a}_i^\perp$ are 2D basis vectors chosen in the form: $\mathbf{a}_i^\perp = (\cos(i 6\pi/5), \sin(i 6\pi/5))$ with $i=0,1,2,3,4$) belong to the acceptance domain [17] which has the form of a regular decagon. The distance between opposite sides of the decagon in the perpendicular space (which determines the size of the acceptance domain) is the projection of the 5D vector <1,-1,-1,1,0>. For the sake of clarity we fix one type of quasilattice by choosing for example $\xi = 3$. Eq. (1) is quite similar to the expression which defines the coordinates of nodes in an ordinary planar periodic lattice: in the lattice all nodes are related by translation vectors which can be expressed as integer linear combinations of only two basic vectors. Analogously, the mathematical object given by Eq. (1), in which all nodes are related by the vectors expressed as integer combinations of *five* basic vectors $\mathbf{a}_i$ in the 2D space, is called a *quasilattice*.

It is useful to note that the PPQ admits ten symmetrically equivalent translations $\mathbf{c}_k$, coinciding with the edges of the corresponding tiles (five pairs of $\mathbf{c}_k$ and -$\mathbf{c}_k$ vectors). These edge translations are not primitive and do not coincide with the vectors $\mathbf{a}_i$. Any $\mathbf{c}_k$ vector is equal to the difference of two $\mathbf{a}_i$ vectors making $4\pi/5$ angle.

In contrast to the action of the usual symmetry elements, quasisymmetry elements relate not all the nodes of the quasilattice but only a part of them. The quasisymmetry axis transforms only the nodes closest to the axis one into another, and is thus called the *local symmetry axis*. The PPQ represents a rare example of the quasilattice in which local axes are *not located* in its nodes. In addition, in some cases quasilattice can have usual *global* point *symmetry* elements, which relate all nodes, independently of their distance to the axis. Namely, in the case $\mathbf{v}=0$ (see Eq. (2)) the PPQ considered in the present work becomes invariant with respect to all global symmetry elements of the $C_{5v}$ point symmetry group.

The density function $\rho(\mathbf{r}) = \sum_j \delta(\mathbf{r} - \mathbf{r}^j)$ of a QC can be expanded in Fourier series of density waves with the number $N$ of basis wave-vectors higher than the dimension $n$ of the space. For a planar pentagonal QC, $N=4$. The free energy of a homogeneous QC is invariant [18] with respect to $N$ zero-energy Goldstone variables. Among them, $n$ linear combinations correspond to the QC displacement $\mathbf{u}$ as a whole $\rho(\mathbf{r}) \rightarrow \rho(\mathbf{r}-\mathbf{u})$, and the remaining $N-n$ combinations (which vanish in the case of a periodic crystal) express in a QC correlated homogeneous mutual shifts ($\mathbf{v}$=constant) of density waves [17, 18]. The elastic free energy of a strained QC is expressed then as an invariant functional of spatial derivatives of two fields $\mathbf{u}$ and $\mathbf{v}$ [18]. A variation of the phason field $\mathbf{v}$ (including the inhomogeneous case) preserves the quasilattice translations and results in the switching of occupied positions [17].